\documentclass[floatfix,5p]{elsarticle}

\usepackage{cuted}
\usepackage{bm}
\usepackage{braket}
\usepackage{amssymb}
\usepackage{graphicx}
\usepackage{dcolumn}
\usepackage{amsmath}
\usepackage{amsfonts}
\usepackage{xcolor}
\usepackage[hidelinks]{hyperref}
\usepackage{mathtools}
\mathtoolsset{showonlyrefs}
\begin{document}

\title{\textit{Ab initio} calculation of the $^3$He$(\alpha,\gamma)^7$Be astrophysical S factor with chiral two- and three-nucleon forces}
%\title{\textit{Ab initio} investigation o $^3$He$(\alpha,\gamma)^7$Be astrophysical S factor with chiral two- and three-nucleon forces}
%\author{M. C. Atkinson$^1$, K. Kravvaris$^1$, S. Quaglioni$^1$, P. Navr\'{a}til$^2$} 
\address[1]{Lawrence Livermore National Laboratory, P.O. Box 808, L-414, Livermore, CA 94551, USA}
\address[2]{TRIUMF, 4004 Wesbrook Mall, Vancouver, British Columbia, V6T 2A3, Canada}
\author[1]{M. C. Atkinson\corref{cor1}}
\ead{atkinson27@llnl.gov}
\cortext[cor1]{Corresponding author}
\author[1]{K. Kravvaris} 
\author[1]{S. Quaglioni} 
\author[2]{P. Navr\'{a}til} 

\date{\today}

%\begin{keyword}
   %Nuclear \sep Theory \sep ab initio \sep Reactions \sep Structure \sep S factor
%\end{keyword}

\begin{abstract}

The $^3$He$(\alpha,\gamma)^7$Be radiative capture reaction plays a key role in the creation of elements in stars as well as in the production of solar neutrinos, the observation of which is one of the main tools to study the properties of our sun. Since accurate experimental measurements of this fusion cross section at solar energies are difficult due to the strong Coulomb repulsion between the reactants, the onus falls on theory to provide a robust means for extrapolating from the region where experimental data is available down to the desired astrophysical regime. 
We present the first microscopic calculations of $^3$He$(\alpha,\gamma)^7$Be with explicit inclusion of three-nucleon forces. Our
prediction of the astrophysical $S$ factor qualitatively agrees with experimental data. We further incorporate experimental bound-state and scattering information in our
calculation to arrive at a more quantitative description. This process reveals that our current model lacks sufficient repulsion in the $1/2^+$ channel of our model space to simultaneously reproduce elastic-scattering data. This deficit suggests that $^3$He$(\alpha,\gamma)^7$Be probes aspects of the nuclear force that are not currently well-constrained. 

%While this calculation results in 
%Our calculation 
%We have found that our calculation lacks sufficient repulsion in the $1/2^+$ channel to reproduce scattering data, which could indicate that this reaction probes aspects of the nuclear force that are not constrained by the typical nucleon-nucleon scattering data. 
%By combining the available bound- and scattering-state data, we provide an accurate prediction of radiative capture.

%Could provide a new probe of the nuclear force. 

\end{abstract}

%demonstrate how we can further enhance our prediction of radiative capture. 

\maketitle

The $^3$He$(\alpha,\gamma)^7$Be radiative capture reaction, where $^3$He nuclei combine with $\alpha$ particles ($^4$He nuclei) to form $^7$Be and emit a photon,  played a crucial role in the formation of the lightest elements during the early phases of the universe. 
The
amount of $^7$Li produced in the first 200 seconds after the big bang predicted by big-bang nucleosynthesis models is highly dependent on the astrophysical $S$ factor of the $^3$He$(\alpha,\gamma)^7$Be reaction~\cite{Burles:1999}. 
In addition to events far in the past, the
$^3$He$(\alpha,\gamma)^7$Be reaction is an important part of ongoing processes occurring in young stars the size of our sun. 
In the second branch of the proton-proton reaction network (pp-II), the
$^3$He$(\alpha,\gamma)^7$Be reaction is key to determining neutrino fluxes resulting from the decay of $^7$Be and $^8$B. In standard solar model (SSM) predictions of these neutrino fluxes, the
$^3$He$(\alpha,\gamma)^7$Be capture rate is the largest source of uncertainty from nuclear input~\cite{Acharya:2024}. 

The importance of this capture reaction has made it the focus of many experiments over multiple decades~\cite{Singh:2004,Confortola:2007,DiLeva:2009,Carmona-Gallardo:2015,Bordeanu:2013}.  
While the abundance of reactants in a stellar environment compensates for the exponential Coulomb suppression of fusion, the $^3$He$(\alpha,\gamma)^7$Be capture cross section at astrophysical energies is exceedingly difficult
to measure in terrestrial settings. Due to this limitation, the lowest-energy $S$ factor measurement is at a center-of-mass energy of 90 keV while the Gamow peak (the energy at which the fusion probability is maximized considering the Maxwellian velocity distribution of nuclei at solar temperatures) is around 18 keV~\cite{Acharya:2024}. Thus, theoretical
calculations are necessary to extrapolate the $S$ factor down to solar energies. There have been numerous theoretical calculations of the $^3$He$(\alpha,\gamma)^7$Be $S$ factor including
external-capture models~\cite{Tombrello:1963}, halo-EFT approaches~\cite{Zhang:2020,Higa:2018}, microscopic approaches ~\cite{Nollett:2001,Neff:2011,Kajino:1986,Csoto:2000}, and $\textit{ab initio}$ approaches~\cite{Dohet-Eraly:2016}.
By combining a subset of these theoretical predictions with a curated set of the $S$ factor measurements, an evaluation of the $S$ factor was performed in Solar Fusion III (SF III) resulting in $S_{34}(0) =
0.561\pm0.018 \textrm{ (exp)} \pm 0.022 \textrm{ (theory)}$ keV~\cite{Acharya:2024}. Not only is the theory uncertainty comparable to that from experiment, but it has grown since the previous evaluation~\cite{Adelberger:2011}. This enduring uncertainty continues to motivate theoretical works toward more accurate predictions of $S_{34}(E)$~\cite{Neff:2011,Dohet-Eraly:2016,Nollett:2007,Zhang:2020,Higa:2018}. 

In this Letter, we apply the \textit{ab
initio} framework of the no-core shell model with continuum (NCSMC)~\cite{Navratil:2016,Simone:2013,Baroni2013L} to describe the $^3$He+$\alpha$ system and calculate the $^3$He($\alpha$,$\gamma$)$^7$Be capture reaction, starting form a many-body chiral Hamiltonian with explicit inclusion of three-nucleon (3N) forces. This work is the first calculation of this reaction to include $3N$ forces, which have been shown to be essential in describing big bang and solar fusion cross sections~\cite{Hebborn:2022, Kravvaris:2023}.
%Furthermore, we present a novel way of combining our results with scattering and bound-state experimental data to arrive at a more accurate prediction of the $S_{34}(E)$ $S$ factor. 
The present results are an important step toward improving the evaluation of the zero-energy $S$ factor, $S_{34}(0)$, and hence reduce the uncertainty of SSM calculations. 

The $^3$He$(\alpha,\gamma)^7$Be radiative-capture cross section at astrophysically relevant energies can be written as~\cite{Descouvemont:2005}
\begin{align}
   \label{eq:capture}
\sigma(E) &= \frac{64\pi^4}{4\pi\epsilon_0\hbar\nu}\sum_{\kappa\lambda}\frac{k_\gamma^{2\lambda+1}}{[(2\lambda+1)!!]^2}\frac{\lambda+1}{\lambda} \\ \nonumber
&\times
\sum_{J_i\ell_is_i}\frac{\hat{J}^2_f}{\hat{s}^2_P\hat{s}^2_T\hat{\ell}^2_i}\left|\Braket{\Psi^{J_f^{\pi_f}T_f}||\mathcal{M}^{\kappa\lambda}||\Psi_{\ell_is_i}^{J_i^{\pi_i}T_i}}\right|^2 ,
\end{align}
where $\Psi^{J_f^{\pi_f}T_f}$ and $\Psi_{\ell_is_i}^{J_i^{\pi_i}T_i}$ correspond to the wave functions for the final bound state ($^7$Be in this case) and the initial scattering state ($^3$He$+^4$He in this case), respectively, with corresponding quantum numbers $J$, $\ell$, $s$, $\pi$, and $T$ representing total angular momentum, orbital angular momentum, spin, parity, and isospin. $s_P$ and $s_T$ are the spin quantum numbers of the projectile
($^3$He) and target ($^4$He) nuclei, respectively, and $\lambda$ is the multipolarity of the electric ($\kappa=E$) and magnetic ($\kappa=M$) transition operators, $\mathcal{M}^{\kappa\lambda}$, the expressions for which can be found in, e.g., Ref.~\cite{Navratil:2016}. We employ the notation where $\hat{s} = \sqrt{2s+1}$. In our calculations, we include electric transition operators up to $\lambda=2$ ($E2$) and magnetic transition operators up to $\lambda=1$ ($M1$). For the energy ranges considered in this radiative-capture calculation, the contribution from the $E1$ transition is dominant while those of $M1$ and $E2$ are almost negligible~\cite{Dohet-Eraly:2016} (and therefore higher order terms are ignored).
The capture cross section can be factored as 
\begin{equation}
   \sigma(E) = \frac{S_{34}(E)}{E}\textrm{exp}\left\{-\frac{2\pi Z_1Z_2e^2}{\hbar\sqrt{2E/m}}\right\},
   %\frac{S_{34}(E)}{E} = \sigma(E)\textrm{exp}\left\{\frac{2\pi Z_1Z_2e^2}{\hbar\sqrt{2E/m}}\right\}.
   %S_{34}(E) = E\sigma(E)\textrm{exp}\left\{\frac{2\pi Z_1Z_2e^2}{\hbar\sqrt{2E/m}}\right\}.
   \label{eq:s34}
\end{equation}
where $S_{34}(E)$ isolates the nuclear component of $\sigma(E)$ and $1/E$ and the exponential term, respectively, account for reaction kinematics and tunneling through the Coulomb barrier.

Both the initial scattering and final bound states in Eq.~\eqref{eq:capture} are calculated within the NCSMC framework through the explicit inclusion of $^3$He+$\alpha$ clustering in the many-body wave function. 
The ansatz for the NCSMC initial and final state is a generalized
cluster expansion~\cite{Navratil:2016}
\begin{align}
   \Ket{\Psi^{J^\pi T}} = &\sum_\lambda c_\lambda^{J^\pi T}\Ket{A \lambda J^\pi T} \nonumber\\
   &+ \sum_\nu\int drr^2\frac{\gamma_{\nu}^{J^\pi T}(r)}{r}\hat{\mathcal{A}}_\nu\Ket{\Phi_{\nu r}^{J^\pi T}}.
   \label{eq:ncsmc}
\end{align}
The first term on the right-hand side of the equation is an expansion over translationally invariant eigenstates, $\Ket{AJ^\pi T}$, of the aggregate system ($^{7}$Be in this case) calculated within the no-core shell model (NCSM)~\cite{Barrett:2013}. The NCSM is an \textit{ab initio} many-body method for the description of static wave functions that allows for the use of both Jacobi relative coordinate ~\cite{Navratil:2000} and single-particle Slater determinant basis states~\cite{Navratil:1998, Navratil:2000b, Navratil:2000c}. The second term is an expansion over fully antisymmetrized microscopic cluster basis channels which describe the $^3$He and $^4$He clusters in relative motion~\cite{Quaglioni:2009,Quaglioni:2008},
\begin{align}
   \Ket{\Phi_{\nu r}^{J^\pi T}} = &\left[\left(\Ket{^4\textrm{He}\lambda_4J_4^{\pi_4}T_4}\Ket{^3\textrm{He}\lambda_3J_3^{\pi_3}T_3}\right)^{(sT)}Y_\ell(\hat{r}_{34})\right]^{(J^{\pi}T)} \nonumber \\
   &\times \frac{\delta(r-r_{34})}{rr_{34}}.
\label{eq:cluster}
\end{align}
Here, $\Ket{^4\textrm{He}\lambda_4J_4^{\pi_4}T_4}$ is an NCSM eigenstate of $^4$He with energy label $\lambda_4$, total angular momentum $J_4$ ($s_T$ in Eq.~\eqref{eq:capture}), parity $\pi_4$, and isospin $T_4$, $\Ket{^3\textrm{He}\lambda_3J_3^{\pi_3}T_3}$ is analogously defined for $^3$He, $s$ is the channel spin, $\bm{r}_{34} = r_{34}\hat{r}_{34}$ is the relative radial coordinate between the centers of mass of $^3$He and $\alpha$, and $\nu$ is a collective index of the relevant quantum numbers.
The projectile and target wave functions are once again described within the NCSM approach. 
%Here $r$ denotes the distance between the clusters and $\nu$ is a collective index of the relevant quantum numbers.

The discrete coefficients, $c_\lambda^{J^\pi T}$, and continuous relative-motion amplitudes, $\gamma_{\nu}^{J^\pi T}(r)$, are obtained as solutions to the coupled equations \cite{Simone:2013,Baroni2013L}
\begin{align}
   \begin{pmatrix}
      H_\textrm{NCSM} & \bar{h} \\
      \bar{h} & \bar{\mathcal{H}}
   \end{pmatrix}
   \begin{pmatrix}
      c \\
      \chi
   \end{pmatrix}
   =
   E
   \begin{pmatrix}
      1 & \bar{g} \\
      \bar{g} & 1
   \end{pmatrix}
   \begin{pmatrix}
      c \\
      \chi
   \end{pmatrix}.
   \label{eq:bloch}
\end{align}
Here, $(H_\textrm{NCSM})_{\lambda\lambda^{'}} = E_\lambda\delta_{\lambda\lambda^{'}}$ is the expectation value of the Hamiltonian in the NCSM model space which evaluates to a diagonal matrix consisting of NCSM eigenvalues; $\bar{\mathcal{H}}_{\nu\nu^{'}} = \left(\mathcal{N}^{-1/2}\mathcal{H}\mathcal{N}^{-1/2}\right)_{\nu\nu^{'}}$ and $\chi_\nu = \left(\mathcal{N}^{1/2}\gamma\right)_\nu$ are the Hamiltonian kernel and relative wave functions, respectively, where $\mathcal{N}_{\nu\nu^{'}}(r,r') = \braket{\Phi_{\nu^{'}r^{'}}^{J^{\pi}T}|\hat{\mathcal{A}}_{\nu^{'}}\hat{\mathcal{A}}_{\nu}|\Phi_{\nu r}^{J^{\pi}T}}$ and 
$\mathcal{H}_{\nu\nu^{'}}(r,r') = \braket{\Phi_{\nu^{'}r^{'}}^{J^{\pi}T}|\hat{\mathcal{A}}_{\nu^{'}}\hat{H}\hat{\mathcal{A}}_{\nu}|\Phi_{\nu r}^{J^{\pi}T}}$; $\bar{g}_{\lambda\nu}(r)$ and $\bar{h}_{\lambda\nu}(r)$ are the overlap and Hamiltonian form factors describing the coupling between the NCSM sector and the cluster sector of the full basis, respectively, proportional to $\braket{A\lambda J^\pi T|\hat{\mathcal{A}}_\lambda|\Phi_{\nu r}^{J^{\pi}T}}$ and $\braket{A\lambda J^\pi T|\hat{H}\hat{\mathcal{A}}_\lambda|\Phi_{\nu r}^{J^{\pi}T}}$. 
The bound states and scattering matrix (and from it any scattering observable) are then obtained by matching the solutions of Eq.~\eqref{eq:bloch} with the known asymptotic behavior of the wave function at large distances by means of the microscopic R-matrix method~\cite{Baye:2010,Simone:2013}.

%Eq.~\eqref{eq:bloch} is solved using an extension of the coupled-channel R-matrix method on a Lagrange mesh~\cite{Baye:2010,Simone:2013}.

The microscopic $A$-nucleon Hamiltonian, $\hat{H}$, adopted in the present work is built on the nucleon-nucleon ($NN$) chiral
interaction at
next-to-next-to-next-to leading order of Ref.~\cite{Entem:2003}, denoted as $NN$-N$^{3}$LO, along with a three-body interaction at next-to-next-to leading order (N$^{2}$LO) with simultaneous local and nonlocal regularization~\cite{Navratil:2007,Gennari:2018,Gysbers2019NatPhys,Soma2020}.
The whole chiral interaction (two- plus three-body) will be referred to as $NN$-N$^{3}$LO$+3N_{\rm lnl}$.
To accelerate the convergence of the NCSMC calculation, we first soften the chiral interaction through
the similarity renormalization group (SRG)
technique~\cite{Wegner1994,Bogner2007,PhysRevC.77.064003,Jurgenson2009}.
We include SRG induced forces up to the three-body level.
To minimize the influence of four- and higher-body induced terms, we adopt the SRG momentum scale of $\lambda_{\mathrm{SRG}}{=}2.0$ fm$^{-1}$~\cite{PhysRevC.103.035801}. Further,
we choose a harmonic oscillator (HO) frequency of $\hbar\Omega = 20$ MeV that minimizes the ground-state energies of the investigated nuclei~\cite{Navratil:2011}. 

The NCSMC model space for the present calculation consists of microscopic seven-body cluster states built from the $0^+$ and $1/2^+$ ground states of
$^4$He and $^3$He, respectively, and the ten lowest positive and negative parity
eigenstates of $^{7}$Be with total angular momentum $J$ ranging from $1/2$ to $7/2$.
The $^3$He (and even $^4$He) reactants can be deformed in the reaction process, in principle requiring the inclusion of excited states of the reactants to take this deformation into account in a microscopic cluster expansion. In the NCSMC, we compensate for the omission of these excited states by including eigenstates of the $^7$Be aggregate system, which help to describe short range correlations~\cite{Quaglioni:2009}. Among the twenty $^7$Be states, three $1/2^+$ and $3/2^-$ states and two $1/2^-$ states are the most relevant channels in this radiative-capture process. All projectile, target, and aggregate states are calculated within the NCSM using the same $\hat{H}$ that generates the Hamiltonian kernel in Eq.~\eqref{eq:bloch}.

\begin{figure}[h]
   \includegraphics[width=\columnwidth]{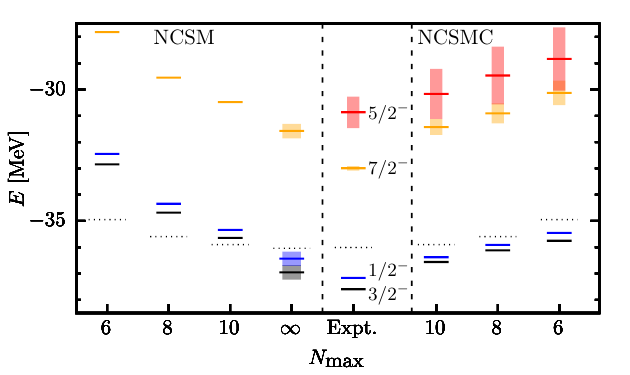}
   \caption{Convergence of calculated energy levels in $^7$Be for increasing basis size $N_\textrm{max}$ compared to experiment~\cite{Tilley:2002}. The horizontal dotted lines correspond to the $^3$He$+\alpha$ threshold at each $N_\textrm{max}$. The left-half of the figure (partitioned by the vertical dashed line) contains the NCSM levels at each $N_\textrm{max}$. The column labeled as ``$\infty$" corresponds to extrapolated energy levels (see text). The right-half of the figure contains NCSMC levels and resonances from solving the scattering equations. Widths of resonant states are represented by shaded regions.}
   \label{fig:levels}
\end{figure}

The HO model space used to compute the NCSMC kernels is $N_{\textrm{max}}=10$ for $^3$He, $^4$He, and negative-parity $^7$Be states and $N_{\textrm{max}}=11$ for positive-parity $^7$Be states. 
To compute the RGM part of the NCSMC kernels, we include matrix elements of the 3N force up to a total number of single-particle quanta for the three-body basis of $E_{3\textrm{max}}=18$.
% The 3N force matrix elements are included up to a total number of single-particle
% quanta for the three-body basis of E3max=16
% The 3N force used to compute the RGM part of the NCSMC kernels was truncated by employing the so-called $E_{3\mathrm{max}}$ truncation scheme CITE, where the sum of HO quanta in each of the particles interacting ($e_i$) is limited to $e_1+e_2+e_3 \leq E_{3\mathrm{max}}$. To accommodate the size of the model space, we generated matrix elements of the 3N force up to e3max 18.
Beyond this mode-space size, the NCSMC calculations become computationally intractable when including $3N$ forces.
In such a model space, the ground-state energies of $^{3}$He
and $^{4}$He are close to experiment ($E_\textrm{expt.}(^3$He$) - E_\textrm{NCSM}(^3$He$) = -0.10$ MeV, $E_\textrm{expt.}(^4$He$) - E_\textrm{NCSM}(^4$He$) = -0.016$ MeV) while the bound states of $^{7}$Be are within 1.5 MeV of experiment. 
Taking advantage of the convergence behavior in $N_\textrm{max}$, we extrapolate our NCSM levels to $N_\textrm{max}\rightarrow\infty$ by using an
exponential decay function $E(N) = Ae^{-bN} + E_\infty$~\cite{More:2013}. The resulting levels (labeled ``$\infty$" in Fig.~\ref{fig:levels}) are closer to experiment and demonstrate that $N_{\textrm{max}}=10$ NCSM eigenstates employed in the following NCSMC calculations are converged within 4\% of their extrapolated values.

%\begin{figure}[h]
   %\includegraphics[width=\columnwidth]{figures/ncsm-converge-n3lo}
   %\includegraphics[width=\columnwidth]{figures/ncsm-nmax}
   %\caption{Convergence of calculated energy levels in $^7$Be for increasing basis size $N_\textrm{max}$ compared to experiment~\cite{Tilley:2002}. The horizontal dashed lines correspond to the $^3$He$+\alpha$ threshold at each $N_\textrm{max}$. (a) The calculated NCSM levels at each $N_\textrm{max}$. The column labeled as "$\infty$`` corresponds to extrapolated energy levels (see text). (b) Calculated NCSMC levels and resonances. Widths of resonant states are represented by shaded regions. The column labeled $10^*$ corresponds to NCSMC calculations using the NCSM levels extrapolated to $N_\textrm{max}\rightarrow\infty$ (see text).}
   %\label{fig:levels}
%\end{figure}

\begin{table}[h]
   \begin{center}
      {\renewcommand{\arraystretch}{1.1}
\begin{tabular}{lcccc} 
\hline
$^7$Be & NCSM & NCSMC & NCSMC$_\textrm{pheno}$ & Expt. \\
\hline
   %$^{3}$He & -7.62 & &  & $-7.72 \pm 5\times10^{-7}$ \\
   %$^{4}$He & -28.28 & & &  $-28.30 \pm 8\times10^{-8}$ \\
   $E_{3/2^-}$ & 0.261 & -0.660 & -1.587 & -1.587\\
   $\mathcal{C}_{3/2^-}$ & - & 2.87 & 3.91 & -\\
   $E_{1/2^-}$ & 0.563 & -0.485 & -1.16 & -1.16\\
   $\mathcal{C}_{1/2^-}$ & - & 2.93 & 3.53 & -\\
   $r_{\textrm{ch}}$ & 2.44 & 2.75 & 2.59 & 2.647(17) \\
   $Q$ & -4.90 & -7.24 & -6.31 & - \\
   $\mu$ & -1.13 & -1.17 & -1.13 & -1.3995(5) \\
   \hline
\end{tabular}
}
\end{center}
\caption{Bound-state properties of $^7$Be generated by the NCSM and NCSMC at $N_\textrm{max}=10$ compared to experimental data~\cite{Nortershauser:2009,Huang:2021,Raghavan:1989} where $E_{3/2^-}$ and $E_{1/2^-}$ are in MeV with respect to the $^3$He$+\alpha$ threshold, $\mathcal{C}_{3/2^-}$ and $\mathcal{C}_{3/2^-}$ are in fm$^{-1/2}$, $r_{ch}$ is in fm, $Q$ is in $e\cdot$fm$^2$, and $\mu$ is in $\mu_N$ (the nuclear magneton). The column labeled ``NCSMC$_\textrm{pheno}$" is the result obtained by phenomenologically adjusting the NCSM eigenvalues employed in the NCSMC to reproduce experimental binding energies (see text).}
\label{tab:energies}
\end{table}

Concerning the NCSMC kernels in Eq.~\eqref{eq:bloch}, the couplings between the microscopic cluster states, $\mathcal{N}_{\nu\nu^{'}}(r,r')$ and $\mathcal{H}_{\nu\nu^{'}}(r,r')$, are calculated using the configuration interaction framework for scattering and reactions induced by light projectiles developed in Ref.~\cite{Kravvaris:2017,Kravvaris:2020cvn}. 
 The couplings between the NCSM sector and the cluster sector, $\bar{g}_{\lambda\nu}(r)$ and $\bar{h}_{\lambda\nu}(r)$, are calculated in a similar way and details will be described in a future publication. 
We calculate the full NCSMC wave function by solving Eq.~\eqref{eq:bloch} using these computed kernels.
The energies computed in the NCSMC are an improvement over the NCSM thanks to the inclusion of the microscopic cluster states of $^3$He$+\alpha$ (see Fig.~\ref{fig:levels} and Table~\ref{tab:energies}).
Especially noteworthy are the $3/2^-$ and $1/2^-$ states which are now bound in the NCSMC. 
The bound levels calculated in the NCSMC change by less than 4\% between $N_{\textrm{max}}=8$ and $N_{\textrm{max}}=10$, demonstrating that the calculation is well-converged. This convergence is further confirmed by the good agreement between the $N_{\textrm{max}}=10$ NCSMC energy levels and the corresponding NCSM values extrapolated to $N_\textrm{max}\rightarrow\infty$. 
Furthermore, the correct description of the wave function at long range allows us to calculate the position and width of the $5/2^-$ and $7/2^-$ resonances which are also changing by less than 4\% from  $N_{\textrm{max}}=8$ to $N_{\textrm{max}}=10$. 
%Additionally, we predict an unobserved $1/2^+$ resonance. Its convergence behavior is unlike the other resonances. The close proximity of this $1/2^+$ resonance to the $p+^6$Li threshold (displayed as a dashed horizontal line in Fig.~\ref{fig:levels}), suggests the resonance has a strong proton component. We attribute the convergence behavior of the $1/2^+$ resonance to the absence of the $p+^6$Li channel in our NCSMC model space.

To probe the radial shape of our $^7$Be ground-state wave function, we calculate the charge radius ($r_{ch}$), electric quadrupole moment ($Q$), and magnetic moment ($\mu$) of the $^7$Be ground state and compare to experiment
(where available) in Table~\ref{tab:energies}. Also included in Table~\ref{tab:energies} are the $^7$Be $3/2^-$ and $1/2^-$ bound-state energies along with their corresponding asymptotic normalization coefficients (ANCs). Just as with the energy levels in Fig.~\ref{fig:levels}, the inclusion of the $\alpha+^3$He cluster state improves the predictions of the NCSMC over the NCSM. The charge radius, in particular, increases by more than 10\% from the NCSM prediction, leaving the NCSMC prediction less than 4\% from the measured values. This improved charge radius calculation is a natural consequence of representing the correct asymptotics through the inclusion of the microscopic cluster states, $\Ket{\Phi_{\nu r}^{J^\pi T}}$, in Eq.~\eqref{eq:ncsmc}~\cite{Quaglioni:2018,Romero:2016}.

%\subsection{Phenomenologically adjusted NCSMC}

\begin{figure}[h]
   \includegraphics[width=\columnwidth]{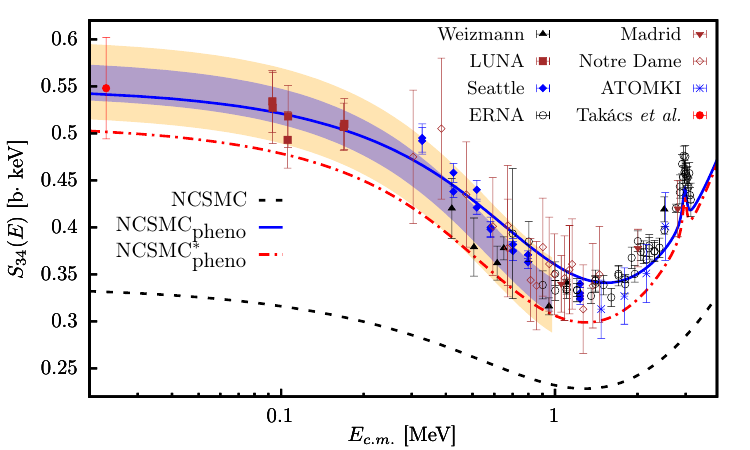}
   \caption{NCSMC calculations of $S_{34}(E)$ at $N_\textrm{max}=10$. The dashed line is the unshifted NCSMC result. The solid line is the NCSMC$_\textrm{pheno}$ result after adjusting to experimental bound-state data. The dot-dashed line is the NCSMC$^*_\textrm{pheno}$ result after adjusting to both experimental bound-state and scattering data.
   The measurements are from Refs.~\cite{Singh:2004}(triangles), ~\cite{Confortola:2007} (squares), ~\cite{Brown:2007} (filled diamonds), ~\cite{DiLeva:2009} (open circles), ~\cite{Carmona-Gallardo:2015} (upside-down triangles), ~\cite{Kontos:2013} (open diamonds), ~\cite{Bordeanu:2013} (stars), and the filled circle at the Gamow peak is deduced from solar-neutrino data~\cite{Takacs:2015}. The shaded region represents the current evaluation from SF III, the form of which is reported as a polynomial fit to halo-EFT results. The widths of the regions are based on the reported uncertainties of $S(0)$ (with different shades representing
      uncertainty from theory or experiment)~\cite{Acharya:2024}.}
   \label{fig:capture-shifted}
\end{figure}

%Using the bound- and scattering-state wave functions calculated in the NCSMC, we employ Eq.~\eqref{eq:capture} to calculate the radiative capture cross section.  As mentioned in Sec.~\ref{sec:intro}, the capture
%cross section at astrophysically-relevant energies is heavily suppressed by the Coulomb repulsion between the fusing nuclei. Therefore, the Coulomb contribution is
%typically factored out of capture cross sections leaving the so-called $S$ factor
%\begin{equation}
   %%\sigma(E) = \frac{S_{34}(E)}{E}\textrm{exp}\left\{-\frac{2\pi Z_1Z_2e^2}{\hbar\sqrt{2E/m}}\right\}.
   %%\frac{S_{34}(E)}{E} = \sigma(E)\textrm{exp}\left\{\frac{2\pi Z_1Z_2e^2}{\hbar\sqrt{2E/m}}\right\}.
   %S_{34}(E) = E\sigma(E)\textrm{exp}\left\{\frac{2\pi Z_1Z_2e^2}{\hbar\sqrt{2E/m}}\right\}.
   %\label{eq:s34}
%\end{equation}
%The $S$ factor is the component of the capture cross section that is informed by nuclear processes.   The subscript in Eq.~\eqref{eq:s34}, $S_{34}(E)$, indicates that
%this is specifically the $S$ factor for the $^3$He$(\alpha,\gamma)^7$Be reaction (3 representing $^3$He and 4 representing $^4$He). 

%The $S$ factor obtained directly from the bound and scattering wave functions 
%The bound- and scattering-state wave functions discussed above 

Using the bound- and scattering-state wave functions calculated in the NCSMC, we employ Eqs.~\eqref{eq:capture} and~\eqref{eq:s34} to calculate $S_{34}(E)$, yielding the dashed line in Fig.~\ref{fig:capture-shifted}, which is in qualitative agreement with the experimental data. While our prediction is below the experimental data, the shape matches that of the data except at higher energies where we miss the peak induced by the $7/2^-$ resonance (which lies at higher energy in our calculation). 
To understand the origin of the difference between our prediction and the latest evaluation, we consider a series of adjustments to the NCSMC Hamiltonian kernels. 

We apply phenomenological shifts to the $^7$Be $1/2^-$, $3/2^-$, and $7/2^-$ NCSM energy eigenvalues in the upper left quadrant of the NCSMC Hamiltonian kernel (left-hand side of Eq.~\eqref{eq:ncsmc}) to exactly reproduce the corresponding experimental energies. We also adjust the $^{4}$He and $^{3}$He energies entering the definition of the total energy in the right-hand side of Eq.~\eqref{eq:bloch} to match experiment even though these minor shifts (see Fig.~\ref{fig:levels}) result in almost negligible changes to the NCSMC results.
These phenomenological corrections improve not only the energies but also the radial shape of the $3/2^-$ and $1/2^-$ bound states, yielding more quantitative predictions for the observables in Table~\ref{tab:energies} and resulting in an increase of the overall normalization of $S_{34}(E)$.
The $S$ factor obtained after these phenomenological shifts, dubbed NCSMC$_\textrm{pheno}$, is shown by 
the solid line in Fig.~\ref{fig:capture-shifted}. The NCSMC$_\textrm{pheno}$ prediction quantitatively reproduces the experimental data and agrees with the SF III evaluation over the entire range of energies between threshold and 4.5 MeV.

\begin{figure*}[h]
   \includegraphics[width=\linewidth]{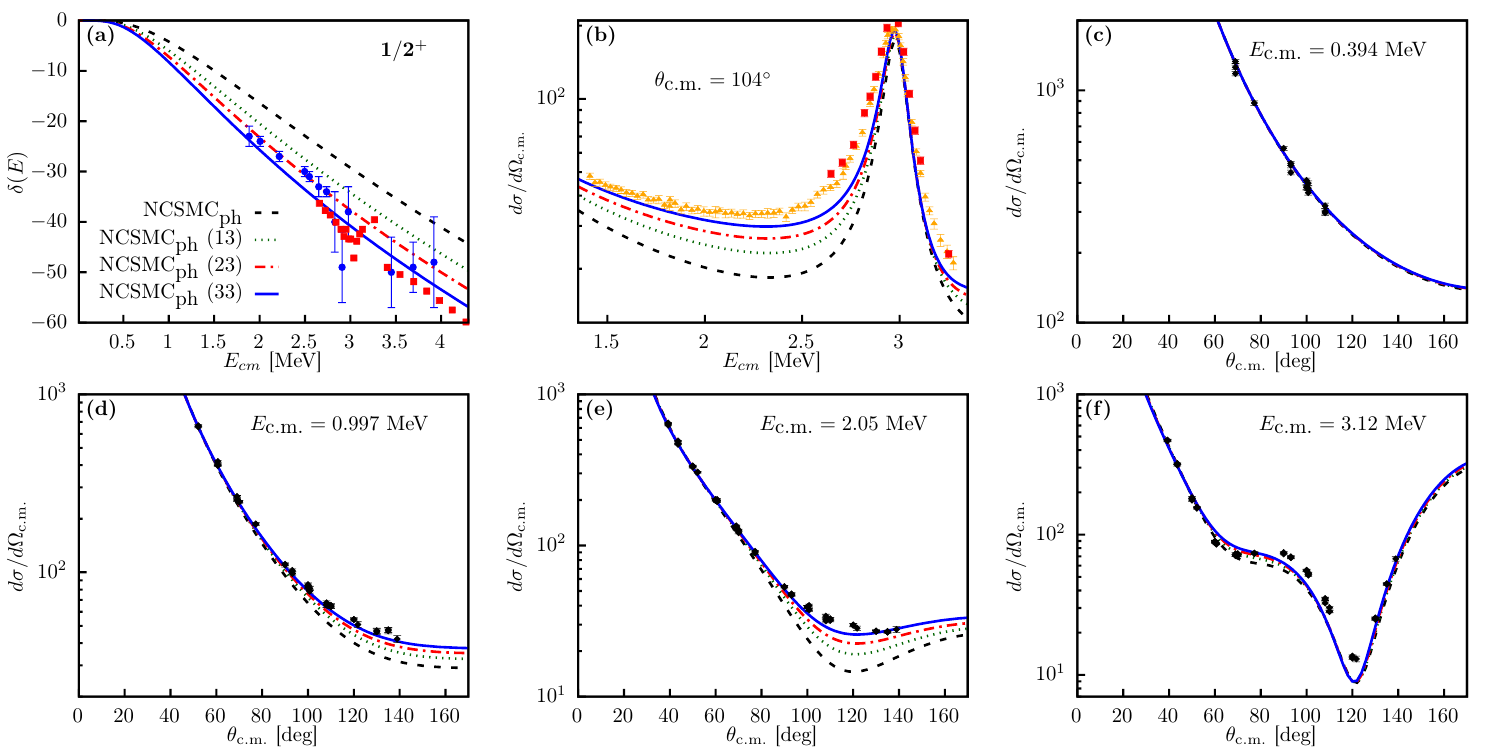}
   \caption{This figure illustrates the effect of adding repulsion to the NCSMC Hamiltonian kernel (see Eq.~\eqref{eq:repulsion}). Each panel contains 4 lines representing NCSMC$_\textrm{pheno}$, and starting from the dashed line, each line has increasing magnitudes of added repulsion. The dashed line has no added repulsion, the dotted line has 13 MeV repulsion added, the dot-dashed line has 23 MeV repulsion added, and the solid line
   has 33 MeV repulsion added. (a) Reduction of the $1/2^+$ phase shift due to added repulsion. The circle points are phase shifts extracted from the elastic-scattering experiment in Ref.~\cite{Hardy:1972}, and the square points are extracted from the elastic-scattering experiment in Ref.~\cite{Spiger:1967}. (b) Increased
   strength in the NCSMC-calculated differential cross section as a function of energy for a fixed $\theta_{\textrm{c.m.}}=104^\circ$ due to added repulsion. The triangle points represent data measured at $\theta_{\textrm{c.m.}}=104^\circ$~\cite{Barnard:1964}. The square points represent data measured at $\theta_{\textrm{c.m.}}=106.4^\circ$~\cite{Spiger:1967}. (c)-(f) Increased strength in the elastic-scattering angular distributions at $E_\textrm{c.m.}=0.394,0.997,2.05,3.125$ MeV due to added
   repulsion. Data from Ref.~\cite{sonik:2022}. }
   \label{fig:phase-multi-3x1}
\end{figure*}

We also probe our predictions of the scattering wave functions by calculating
$^3$He$+^4$He elastic-scattering differential cross sections and comparing them with measurements from Refs.~\cite{sonik:2022,Barnard:1964,Spiger:1967}.
Of particular interest is the experiment performed by the SONIK collaboration in 2022 that reaches energies as low as $E_\textrm{lab}=239$ keV~\cite{sonik:2022} and provides better angular range than previous measurements.
The NCSMC is in good agreement with the lower-energy SONIK data (dashed line in panel (c) of Fig.~\ref{fig:phase-multi-3x1}), but the agreement declines at
higher-energy backward angles (dashed line in panels (d) and (e) of Fig.~\ref{fig:phase-multi-3x1}).
The discrepancy at backward angle is also present when comparing to older experimental data sets at
   $\theta_{\textrm{c.m.}}=104^\circ$ and $\theta_{\textrm{c.m.}}=106.4^\circ$\cite{Barnard:1964,Spiger:1967} (dashed line in panel (b) of Fig.~\ref{fig:phase-multi-3x1}). 
   
We note that the logarithmic scale amplifies the differences at large angles owing to the divergence of the Rutherford cross section at $\theta_{\textrm{c.m.}}=0^\circ$. In fact, we find that the difference between the NCSMC results and the SONIK data is a constant (angle-independent) shift of about 10 mb.
A constant shift such as this must be rooted in the s-wave channel of the scattering wave function corresponding to the $1/2^+$ $^3$He$+\alpha$ phase shift. Indeed, the NCSMC phase shift is less-repulsive than the data of Refs.~\cite{Spiger:1967,Tombrello:1963} (see the dashed line in panel (a) of Fig.~\ref{fig:phase-multi-3x1}), indicating that the $1/2^+$ channel of the NCSMC Hamiltonian kernel lacks sufficient repulsion.
Similarly, we find that the $1/2^+$ scattering length, $a_0 = 11.2$ fm,
calculated by fitting the effective range expansion to our
corresponding phase shift, is much smaller than the one derived from an $R$-matrix fit to the SONIK elastic data~\cite{sonik:2022}. This further confirms the lack of repulsion in our $1/2^+$ channel.

\begin{table}[h]
   \begin{center}
\begin{tabular}{cc} 
\hline
Model & $S_{34}(0)$ [keV b] \\
\hline
\hline
NCSMC$_\textrm{pheno}$   & $0.545\pm 0.001$ \\
NCSMC$^*_\textrm{pheno}$   & $0.505$ \\
halo EFT~\cite{Zhang:2020}  & $0.577^{+0.015}_{-0.016}$\\
halo EFT~\cite{Higa:2018}   & $0.558\pm0.008\textrm{(fit)}\pm0.056\textrm{(EFT)}$\\
$R$-matrix~\cite{Odell:2022}   & $0.539^{+0.011}_{-0.012}$ \\
SF III~\cite{Acharya:2024} & $0.561\pm0.018\textrm{(exp)} \pm 0.022\textrm{(th)} $ \\
\hline
% Neff  ~\cite{Neff:2011} (FMD) & - & - & - & $0.593$\\
%Literature & Odell & Zhang & $R$-matrix & Solar Fusion II  \\
%\hline
 %$S_{34}(0)$ & $0.495\pm 0.008$ & $0.578\pm$ & $0.542\pm$ & $0.56\pm0.02 \pm 0.02$ \\
%\hline
\end{tabular}
\end{center}
\caption{$S$ factor of $^3$He$(\alpha,\gamma)^7$Be at zero energy, $S_{34}(0)$. 
The NCSMC$_\textrm{pheno}$ result is after adjusting to experimental bound-state data. The NCSMC$^*_\textrm{pheno}$ result is after adjusting to both experimental bound-state and scattering data.
The quoted $R$-matrix value is specifically excluding the Barnard data set (see Ref.~\cite{Odell:2022} for detail).}
\label{tab:s0}
\end{table}

It is unlikely
that the missing repulsion in the computed $1/2^+$ channel  could be fully explained by a NCSMC calculation at the next largest HO model space dimension of $N_{\rm max}=12/13$ (currently computationally out of reach). As discussed, we obtain well-converged bound state energies as well as resonance centroids and widths (see Fig.~\ref{fig:levels}). Furthermore, the difference in the $1/2^+$ $a_0$ between $N_\textrm{max}=10$ and $N_\textrm{max}=8$ is 1.7 fm as opposed to the difference of over 15 fm when compared to the $R$ matrix fit of the SONIK data~\cite{sonik:2022}.  %We then look to the $NN$ and $3N$ interactions. 
Additionally, the NCSMC investigation of $S_{34}(E)$ in Ref.~\cite{Dohet-Eraly:2016}, which reached the $N_{\rm max}=12/13$ HO model space size using the two-nucleon component of the SRG-evolved $NN$ interaction ($NN$-only), manifests a similar lack of repulsion in the
$1/2^+$ channel. This was originally attributed to the lack of explicit $3N$ forces in the calculation (see Fig. (3) of Ref.~\cite{Dohet-Eraly:2016}). 
While the present study shows that the inclusion of $3N$ forces does introduce additional repulsion, it is not sufficient to reproduce experimental data.

%Additionally, the $NN$-only calculation of Ref.~\cite{Dohet-Eraly:2016} was performed using the same $NN$ interaction as the present study albeit with a slightly
%different SRG momentum resolution scale of $2.15$ fm$^{-1}$, indicating that this discrepancy appears regardless of the chosen $\lambda_\textrm{SRG}$ value.
%Given these observations, we hypothesize that the discrepancy may be related to the $NN$ interaction, the testing of which (through the analysis of results obtained with additional $NN$ interactions) is outside the scope of the current work. 
%Therefore, we hypothesize that the discrepancy may be related to the $NN$ interaction. Testing of such a hypothesis through the analysis of additional $NN$ interaction is outside the scope of the current work.
%Due to the
%computational complexity of these calculations, the analysis of additional $NN$ interactions is outside the scope of the current work.

%[NOT SURE I WANT TO INCLUDE THIS]
%Finally, we performed a calculation using different $NN$ and $3N$ interactions ($NN$n4lo + 3NlnlE7) and found the same lack of repulsion in the $1/2^+$ channel.

For the time being, we emulate the effect of a more repulsive nuclear force in
the $1/2^+$ channel by including a nonlocal Woods-Saxon potential, $V(r,r')$, in the microscopic-cluster quadrant of the Hamiltonian kernel (in the
bottom-right element of the matrix on the left-hand-side of Eq.~\eqref{eq:bloch}) such that
\begin{align}
   \mathcal{H}_{\nu\nu^{'}}^{1/2^+}(r,r') \rightarrow \mathcal{H}_{\nu\nu^{'}}^{1/2^+}(r,r') + V(r,r'),
\end{align}
where
\begin{align}
   \label{eq:repulsion}
   V(r,r') = \frac{V_{ws}}{1+e^{(R-r_{ws})/a_{ws}}} \times e^{(r-r')^2/a_{ws}^2}.
\end{align}
Here $R=\frac{r+r'}{2}$, the width and radius $r_{ws}$ and $a_{ws}$ of the Wood-Saxon potential are fixed at $0.6$ fm to resemble the nonlocality and range of $\mathcal{H}_{\nu\nu^{'}}^{1/2^+}(r,r')$, and $V_{ws}$ is the strength of the repulsion which we varied. As expected, the computed phase shift becomes closer to the empirical data as we increase $V_{ws}$ from 13 to 33 MeV (see the progression from dashed to solid lines in Fig.~\ref{fig:phase-multi-3x1}(a). 
Correspondingly, we observe progressively improved agreement with experiment in the elastic-scattering differential cross sections (see panels (b)-(f) of Fig.~\ref{fig:phase-multi-3x1}). %The results in Fig.~\ref{fig:phase-multi-3x1} 
This indicates that, with suitable repulsion in the $1/2^+$
channel, our prediction of the scattering wave function is realistic and can reproduce experimental elastic data. %We note that
   %Additional
   %repulsion could be added to match the Barnard elastic scattering data at the cost of overestimating the SONIK data. This, in turn, would decrease the absolute normalization of our estimated $S_{34}(E)$ similar to the behavior observed in Ref.~\cite{Odell:2022}.

   The $S$ factor obtained after this series of adjustments to the NCSMC Hamiltonian kernel (dubbed NCSMC$^*_\textrm{pheno}$) still quantitatively reproduces the majority of the experimental data at the bottom of their uncertainties over the entire range of energies, yet it lies below the SF III evaluation for most of the energy range (dot-dashed line in Fig.~\ref{fig:capture-shifted}). It is curious that reproducing the elastic-scattering data from the SONIK collaboration causes our $S$ factor to decrease away from the SF III evaluation. This discrepancy seems to reveal a mild tension between some of the $S$ factor measurements (along with the SF III evaluation) and elastic-scattering measurements, similar to what is observed in Ref.~\cite{Odell:2022}. 
   Indeed, adding more repulsion to match the Barnard elastic scattering data would not only result in overestimating the SONIK data, but it would further reduce the NCSMC$^*_\textrm{pheno}$ prediction in Fig.~\ref{fig:capture-shifted} (again, similar to what is observed in Ref.~\cite{Odell:2022}). 
   Even among the $S$ factor data, the NCSMC$^*_\textrm{pheno}$ result points to a mild tension between the Weizmann~\cite{Singh:2004} and Seattle~\cite{Brown:2007} data sets.

\begin{figure}[h]
   \includegraphics[width=\columnwidth]{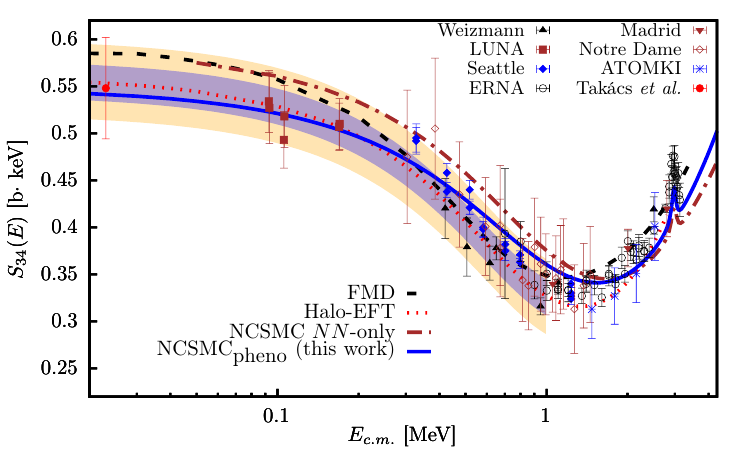}
   \caption{The $S$ factor for the radiative capture process $^3$He$(\alpha,\gamma)^7$Be. The solid line is the present NCSMC$_\textrm{pheno}$ result after adjusting to the bound-state data (see text).
      The dot-dashed line is the previous NCSMC result employing $NN$-only forces~\cite{Dohet-Eraly:2016}. The dashed line is the result of the microscopic FMD calculation from Ref.~\cite{Neff:2011}. The dotted line is a halo-EFT calculation with the updated data-constraints detailed in SF III~\cite{Acharya:2024}. The measurements are from Refs.~\cite{Singh:2004}(triangles), ~\cite{Confortola:2007} (squares), ~\cite{Brown:2007} (filled diamonds), ~\cite{DiLeva:2009} (open circles), ~\cite{Carmona-Gallardo:2015} (upside-down triangles), ~\cite{Kontos:2013} (open diamonds), ~\cite{Bordeanu:2013} (stars), and the filled circle at the Gamow peak is deduced from solar-neutrino data~\cite{Takacs:2015}. The shaded region represents the current evaluation from SF III, the form of which is reported as a polynomial fit to halo-EFT results. The widths of the regions are based on the reported uncertainties of $S(0)$ (with different shades representing uncertainty from theory or experiment)~\cite{Acharya:2024}.} 
      \label{fig:sfactor}
\end{figure}

When compared to the halo-EFT results in Fig.~\ref{fig:sfactor} (which the SF III evaluation is based on), our NCSMC$_\textrm{pheno}$ results match at lower energies and favor a slightly larger $S(E)$ at higher energies.
As shown by the difference between the $NN$-only calculation (dash-dotted line) and the present result, the inclusion of $3N$ forces lowers the absolute normalization of the $S$ factor and molds its shape to be more consistent with the trend of the experimental data. 
The present results differ in both normalization and shape from  those obtained in Ref.~\cite{Neff:2011} (dotted line) using the fermionic molecular dynamics (FMD) approach over the entire energy range.
 % The present results are not dissimilar from those obtained in Ref.~\cite{Neff:2011} (dotted line) using the fermionic molecular dynamics (FMD) approach except at very low energies and above 1 MeV, where the FMD favors somewhat higher $S$ factor values. 
% The situation is similar when comparing with the halo effective field theory (halo-EFT) results of Refs.~\cite{Higa:2018,Acharya:2024}. 

At zero energy, the NCSMC$_{\rm pheno}$ %this results in 
yields $S_{34}(0) = 0.545(1)$ keV b (see Table~\ref{tab:s0}),
where the uncertainty %we consider the uncertainty due to the convergence of the NCSMC predictions with respect to the HO model space size $N_\textrm{max}$. This uncertainty is calculated by taking 
is estimated as the difference between the NCSMC$_\textrm{pheno}$ $N_\textrm{max}=8$ and $N_\textrm{max}=10$ $S_{34}(0)$ predictions. The NCSMC$_\textrm{pheno}$ and NCSMC$^*_\textrm{pheno}$ results in Table~\ref{tab:s0} provide a lower and upper bound for our $S_{34}(0)$ prediction of 0.505 and 0.545 keV b, respectively, that acknowledges the lack of repulsion in our $1/2^+$ channel and the mild tension between elastic-scattering and $S$ factor measurements.    
The halo-EFT values quoted in Table~\ref{tab:s0} are obtained using the experimental data sets detailed in Ref.~\cite{Zhang:2020,Higa:2018}, while the SF III value is obtained using the same halo-EFT model of Ref.~\cite{Higa:2018} but employing an adjusted data set as detailed in Ref.~\cite{Acharya:2024}. 

%\begin{figure}[h]
   %\label{fig:sfactor-nmax}
   %\includegraphics[width=\columnwidth]{figures/capture-nmax}
   %\caption{Convergence in Nmax}
%\end{figure}

In conclusion,
we have presented the first $\textit{ab initio}$ calculation of the $^3$He$(\alpha,\gamma)^7$Be S factor including the 3N force. %$S_{34}(E)$ using a chiral NN+3N Hamiltonian.
%including the $3N$ interaction. 
We demonstrated that 
%can be used not only to make 
the $\textit{ab initio}$ calculations of capture reactions obtained with the NCSMC %of , but also as a framework for incorporating 
can be improved through the incorporation of the information provided by bound-state and scattering measurements
and still maintain predictive capability. 
%to enhance
%predictions of more complex reactions.
While the phenomenological shift applied to reproduce the observed energy levels in $^7$Be
%experimental $^7$Be energies 
can be interpreted as emulating the effect of an infinitely large NCSM model space (see Fig.~\ref{fig:levels}), the repulsive interaction added in the $1/2^+$ channel has no analogous explanation. 
%This reveals that the key to accurately describing scattering observables in $^3$He$+\alpha$ lies in determining what causes the lack of repulsion in the $1/2^+$ channel of the NCSMC scattering wave function. 

Understanding the origin of this lack of repulsion in the $1/2^+$ partial wave will play an important role in accurately describing the $^3$He+$^4$He elastic scattering and $^3$He$(\alpha,\gamma)^7$Be capture cross sections simultaneously. 
Barring the existence of a broad $1/2^+$ resonance which could manifest with the inclusion of the $p+^6$Li channel in our model,
we hypothesize that the present $A=7$ reaction observables probe aspects of chiral interactions that are currently not well-constrained. 
%However, we must also consider the absence of the $p+^6$Li channel in our model space as a potential source for the discrepancy observed between the NCSMC-predicted and empirical phase shifts in the $1/2^+$ channel. 
%The close proximity of the predicted $1/2^+$ resonance to the $p+^6$Li channel (see Fig.~\ref{fig:levels}) indicates that this resonance has strong proton content, and the inclusion of this proton channel could have a non-negligible impact on the $1/2^+$ component of the NCSMC
%scattering wave function. 
Looking ahead, we plan to include the $p+^6$Li channel as well as conduct a systematic analysis of additional chiral interaction models to explore this puzzle in the $1/2^+$ channel.
%We plan to explore these concepts in a future work to determine what causes
%the lack of repulsion in the $1/2^+$ channel.
The present calculation also sets the stage for a more extensive study with different chiral interactions (at several orders) and exploiting correlations among the measured S factor data to arrive at an accurate evaluation with reduced uncertainty in a manner similar to that of Ref.~\cite{Kravvaris:2023}.

The authors would like to thank Gautam Rupak for sharing his halo-EFT $S$ factor results.
We also thank Chlo{\"e} Hebborn for helpful discussions. 
Computing support for this work came from the Lawrence Livermore National Laboratory (LLNL) institutional Computing Grand
Challenge program and from an INCITE Award on the Summit and Frontier supercomputers of the Oak Ridge Leadership Computing Facility (OLCF) at ORNL. 
This material is based upon work supported by the U.S.
Department of Energy, Office of Science, Office of Nuclear Physics, under Work Proposal No. SCW0498.
This work was performed under the auspices of the U.S. Department
of Energy by LLNL under contract DE-AC52-07NA27344.
PN acknowledges support from the NSERC Grant No. SAPIN-2022-00019. 
TRIUMF receives federal funding via a contribution agreement with the National Research Council of Canada. 

\bibliographystyle{apsrev4-1}
\bibliography{paper}

\end{document}